\documentclass[conference]{IEEEtran}
\IEEEoverridecommandlockouts
\usepackage{cite}
\usepackage{amsmath,amssymb,amsfonts}
\usepackage{algpseudocode}
\usepackage{algorithm}
\usepackage{graphicx}
\usepackage{textcomp}
\usepackage{xcolor}
\usepackage{geometry}
\usepackage{subcaption}
\usepackage{booktabs}
\usepackage{bm} 
\usepackage{amsmath, amsthm}
\usepackage{lipsum}
\usepackage{tabularx}
\usepackage{booktabs}
\usepackage{makecell}
\def\BibTeX{{\rm B\kern-.05em{\sc i\kern-.025em b}\kern-.08em
    T\kern-.1667em\lower.7ex\hbox{E}\kern-.125emX}}
\begin{document}

\title{Towards a Partial Computation offloading in In-networking Computing-Assisted MEC: A Digital Twin Approach}

\author{
    \IEEEauthorblockN{Ibrahim Aliyu\textsuperscript{\dag}, Awwal Arigi\textsuperscript{\ddag}, Seungmin Oh\textsuperscript{\dag}, Tai-Won Um\textsuperscript{\S}, and Jinsul Kim\textsuperscript{\dag}}
    \IEEEauthorblockA{
        \textsuperscript{\dag}Department of ICT Convergence System Engineering, Chonnam National University, Gwangju 61186, Korea\\
        \textsuperscript{\ddag}Department of Humans and Automation, Institute for Energy Technology, 1777 Halden, Norway\\
        \textsuperscript{\S}Graduate School of Data Science, Chonnam National University, Gwangju 61186, Korea\\
    }
  \IEEEauthorblockA{  
    Corresponding Authors: Tai-Won Um (email: stwum@jnu.ac.kr) and Jinsul Kim (email: jsworld@jnu.ac.kr).
    }
}
\maketitle

\begin{abstract}
This paper addresses the problem of minimizing latency with partial computation offloading within Industrial Internet-of-Things (IoT) systems in in-network computing (COIN)-assisted Multiaccess Edge Computing (C-MEC) via ultra-reliable and low latency communications (URLLC) links. We propose a digital twin (DT) scheme for a multiuser scenario, allowing collaborative partial task offloading from user equipment (UE) to COIN-aided nodes or MEC. Specifically, we formulate the problem as joint task offloading decision, ratio and resource allocation. We employ game theory to create a low-complexity distributed offloading scheme in which the task offloading decision problem is modelled as an exact potential game. Double Deep Q-Network (DDQN) is utilized within the game to proactively predict optimal offloading ratio and resource allocation. This approach optimizes resource allocation across the whole system and enhances the robustness of the computing framework, ensuring efficient execution of computation-intensive services. Additionally, it addresses centralized approaches and UE resource contention issues, thus ensuring faster and more reliable communication.

\end{abstract}


\section{Introduction}
The convergence of advancements in communication, artificial intelligence, and robust computing architecture is driving the development of a wide range of computation-intensive and time-sensitive services. Multiaccess Edge Computing (MEC) has emerged as a key solution, facilitating remote offloading for such services. However, MEC often faces limitations and security concerns, making it challenging to accommodate the demands of numerous users \cite{chen2022dynamic}.

In contrast, the COIN paradigm, aimed at minimizing latency and improving Quality of Experience (QoE), efficiently utilizes untapped network resources for specific tasks \cite{aliyu2023toward}. However, integrating additional computing resources or enabling in-network computing may escalate power consumption in the network, introducing a trade-off between time delay and energy consumption. Considering the coexistence of COIN with existing edge computing solutions, exploring partial subtask offloading in collaborative scenarios becomes crucial.

The emergence of Digital Twinning, a key concept in the metaverse replicating physical objects and environments, is gaining traction in various domains, including communication networks. Integrating Digital Twinning into edge computing opens exciting possibilities for transforming resource allocation in terms of intelligence, efficiency, and flexibility \cite{wu2021digital}. 
Recent studies have primarily concentrated on DT-assisted task offloading in MEC \cite{do2022digital,van2022edge, li2022digital,hao2023digital}. For example, a study \cite{liu2021digital} addresses the DT-assisted task offloading problem, including mobile-edge server selection for optimizing computing overhead using DDQN in an edge collaboration scenario. Another work \cite{li2022digital} focuses on energy optimization in MEC using DDQN, while \cite{hao2023digital} employs combinatorial optimization to tackle computing overhead in MEC. However, these studies mainly address binary offloading, a critical oversight for the Metaverse. In the Metaverse, tasks often comprise multiple subtasks that can be distributed and processed across various computing nodes, such as COIN nodes and MEC. To fully leverage the advantages of COIN, it's crucial to explore partial offloading. This approach enables COIN and MEC to collaboratively manage compute-intensive tasks by handling multiple subtasks efficiently.

 Motivated by the aforementioned limitation, this paper introduces a DT-aided C-MEC architecture that provides network resources for computation-intensive services. The DT is utilized to model the computing capacity of in-network computing-enabled nodes and edge servers, optimizing resource allocation across the entire system. The main contributions of this paper are as follows. First, we formulate the system utility maximization problem that jointly optimizes the offloading decision, offloading ratio, and resource allocation. Secondly, a distributed game-theoretic approach is proposed for partial computation offloading decision as an exact potential game (EPG) with Nash equilibrium (NE). Within the game, we employed DDQN to predict the future offloading ratio and resource allocation. Finally, our evaluation demonstrates that, in different scenarios, our proposed scheme consistently enhances system utility compared to baselines. It systematically optimizes resource allocation across the whole system and enhances the robustness of the computing framework, ensuring efficient execution of computation-intensive services. Moreover, it effectively tackles centralized approaches and UE resource contention concerns, ensuring accelerated and reliable communication.

\section{System model and problem formulation}

\begin{figure}[ht]
  \centering
  \includegraphics[width=0.5\textwidth]{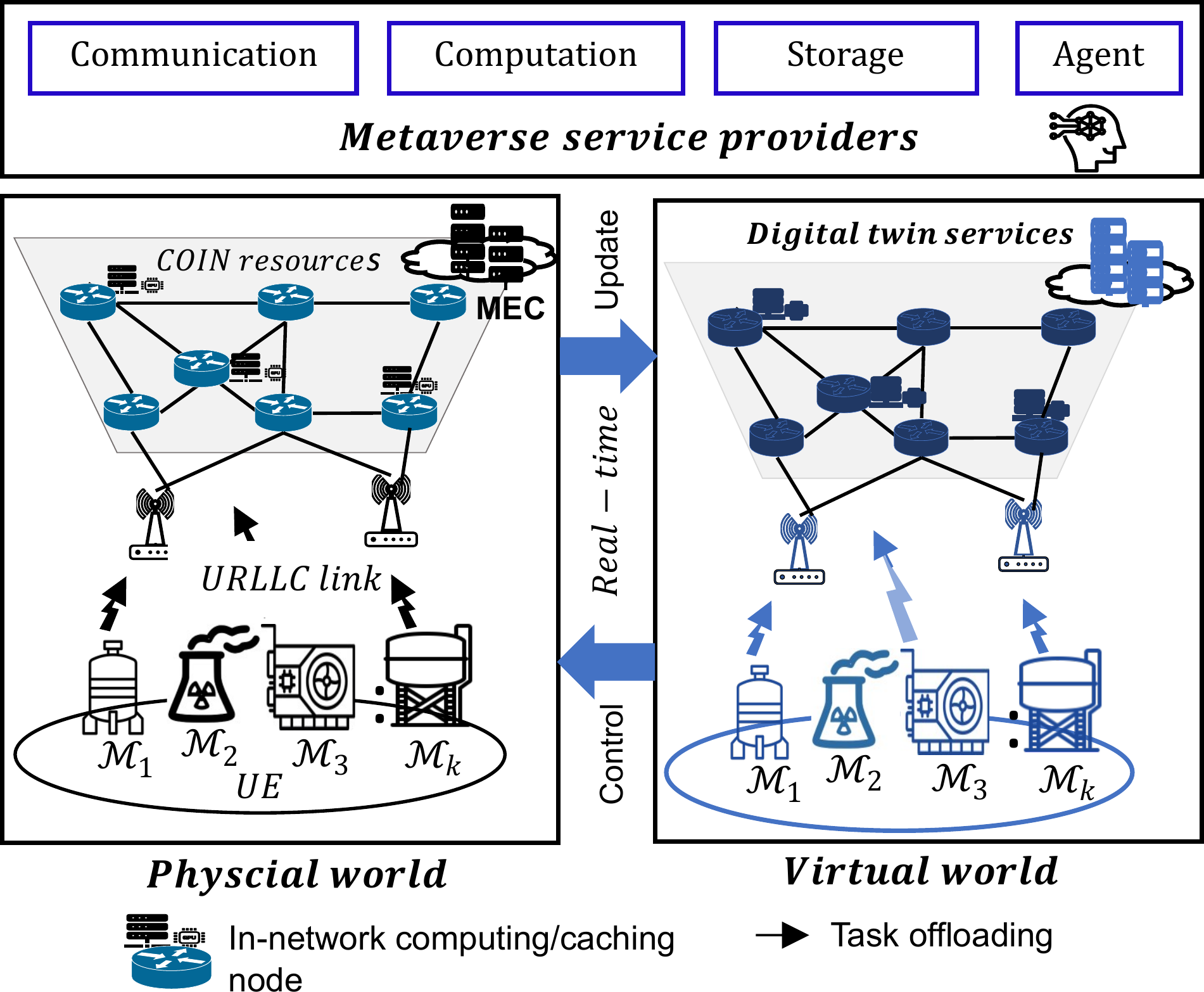}
  \caption{C-MEC architecture}
  \label{C-Mec architecture}
\end{figure}

The C-MEC network architecture system model is illustrated in Fig.\ref{C-Mec architecture}. The model consists of a physical layer which consists of user equipment (UE) and network resources such as COIN-enabled computing nodes (CNs) and MEC servers (ESs) at the edge. This network infrastructure supports the operation of DT services by optimizing resource allocation and enables the whole system via a real-time interaction mechanism. 

Let $\mathcal{M} = \{1, 2, \ldots, M\}$ be the set of $M$ user equipments (UEs), $\mathcal{K} = \{1, 2, \ldots, K\}$ be the set of $K$ COIN computing nodes (CNs), and $R$ be the ES. The CNs and ESs are associated with an access point (AP) to connect the UEs. To ensure high-reliable performance and low latency in the IoT, URLLC short packet communication is employed between the UEs and APs.The system model is as follows:

\subsubsection{Offloading Model in C-MEC Network}
Considering a time slot model, the UEs and CNs are fixed within each time and vary over different time slots. At each time slot $t$, each UE has a computational task characterized by $J_m = \{\eta_m, T_m^{\max}\}$ where $\eta_m = \frac{C_m}{I_m}$ is the task complexity (cycle/bits), $I_m$ is the task size in bits, $C_m$ is the required CPU cycles (cycles) to execute the task, and $T_m^{\max}$ is the maximum tolerable latency for the task $J_m$.

In our scenario, we focus on partial offloading to utilize parallel processing for latency reduction. For instance, in real-time digital twinning of the physical world, numerous devices/sensors collect various views/scenes to reconstruct them in 3D, as discussed in \cite{yu2023asynchronous}. This process of converting the 2D physical world into 3D models necessitates partial computation for enhanced efficiency. Thus, tasks can be subdivided into ratios: one ratio executed by the CNs and the other part executed by the ES, indexed by 0. The ESs can serve multiple UEs while the CN is limited.

Let $\Phi_P = \{\lambda_{mk}, \aleph_m\}$ be the offloading ratio variable where $\lambda_{mk}$ is the portion executed at the CNs, and $\aleph_m = 1 - \lambda_m$ is the portion of the task executed at the ES. Offloading resources are indicated by the variable $\Phi_L = \{\Phi_{\lambda}, \Phi_{\aleph}\}$ where $\Phi_{\lambda}$ and $\Phi_{\aleph}$ indicate tasks execution resource (location) at the CNs and ES, respectively. We assume tasks are generated with high granularity, enabling partial offloading ability. For the task $J_m$, $I_m = \aleph_m I_m + \sum_{k\in K} \lambda_{mk} I_m$ and $C_m = \aleph_m C_m + \sum_{k\in K} \lambda_{mk} C_m$ satisfy $\aleph_m + \sum_{k\in K} \lambda_{mk} = 1$.

\subsubsection{C-MEC DT Model}
DT services generate virtual replicas of physical systems, replicating hardware, applications, and real-time data. The URLLC-based C-MEC's DT is defined as
$\ DT = \{\tilde{\mathcal{M}},\ \tilde{\Phi}_L\},$ where $\{\tilde{\mathcal{M}},\ \tilde{\Phi}_L\}$ represents the system's virtual mirror, including $M$ UEs and $\tilde{\Phi}_L$ C-MEC computing resources (CNs and ES). The DT layer, informed in real-time, automates control via services like data analysis, decision-making, and instant optimization, focusing on tasks like offloading strategies and resource allocation.

Each $m$-th UE's specific DT is associated with a CN node for processing and defined as $\ DT_m^{cn} = (f_m^{cn}, \tilde{f}_m^{cn}),$ where $f_m^{cn}$ denotes the estimated processing rate, while $\tilde{f}_m^{cn}$ quantifies the variation from the actual processing rate between the physical UE and its DT \cite{van2022urllc}. In the DT layer, the critical estimated processing rate, $f_m^{cn}$, mirrors UE behaviors, driving optimization decisions for device configurations. This rate is the focus of our optimization, with its deviation set as a predetermined percentage for simulations, following established practices \cite{do2022digital}.

Likewise, for the $\tilde{\Phi}_L$-th C-MEC computing resource (CNs and ES), its DT ($DT_{\tilde{\Phi}_L}^{cn}$) is formulated as $DT_{\tilde{\Phi}_L}^{cn} = (f_{\tilde{\Phi}_L}^{cm}, \tilde{f}_{\tilde{\Phi}_L}^{cm}),$ where $f_{\tilde{\Phi}_L}^{cm}$ signifies the estimated processing rate of the real C-MEC, and $\tilde{f}_{\tilde{\Phi}_L}^{cm}$ characterizes the disparity in processing rate estimation when compared to the actual C-MEC. The DT emulation of C-MEC (CNs and ES) provides valuable insights into C-MEC processing rates, facilitating efficient allocation of computing resources and reducing processing latency through offloading ratio and computing resource allocation adjustments.

\subsection{Communication Model}
The AP, with $L$ antennas serving $M$ single-antenna UEs, establishes channel connections with compute resource $\tilde{\Phi}_L$ represented by $\bm{h}_{m\tilde{\Phi}_L} = \sqrt{g_{m\tilde{\Phi}_L}} \cdot \bm{ \bar{h}}_{m\tilde{\Phi}_L}$, where $g_m$ is the large-scale channel coefficient and $\bm{ \bar{h}}_{m\tilde{\Phi}_L}$ is small-scale fading following $CN(0, \bm{I})$, where $CN(.,.)$ represents a complex circularly symmetric Gaussian distribution. A channel matrix $\bm{H}_{\tilde{\Phi}_L} = [\bm{h}_{1\tilde{\Phi}_L}, \bm{h}_{2\tilde{\Phi}_L}, \ldots, \bm{h}_{M\tilde{\Phi}_L}] \in \mathbb{C}^{L \times M}$ contains connections from $m$-th SM to the $\tilde{\Phi}_L$-th AP. Each UE's allocated bandwidth, $b_m$. Match filtering and successive interference cancellation (MF-SIC) is employed to improve transmission performance \cite{fang2001performance}. Then, the signal-to-interference-plus-noise (SINR) at the $\tilde{\Phi}_L$-th AP by the $m$-th UE is defined as $\gamma_{m\tilde{\Phi}_L}(p,n) = \frac{p_{m\tilde{\Phi}_L} \lVert h_{m\tilde{\Phi}_L} \rVert^2}{T_{m\tilde{\Phi}_L}(p,n) + N_0},$
where $p_{m\tilde{\Phi}_L}$ is the transmit power of the $m$-th UE, $N_0$ is the noise power, $p = [p_{m\tilde{\Phi}_L}]_{m=1}^M$, and $\mathcal{I}_{m\tilde{\Phi}_L}(p,n) = \sum_{n>m}^{M} p_{n\Phi_L} \frac{\lvert h_{m\Phi_L}^H h_{n\Phi_L} \rvert^2}{\lVert h_{m\Phi_L} \rVert^2} $ is the interference imposed by UEs $n > m$.
Thus, the uplink URLLC transmission rate is expressed as \cite{ren2020joint,she2017radio}:
\begin{equation}
\begin{split}
    \omega_{m\Phi_L}(\bm{p}, \bm{n}) &\approx B \log_2\left[1 + \gamma_{m\Phi_L}(\bm{p}, \bm{n})\right] \\
    &\quad - B\sqrt{\frac{V_{m\Phi_L}(\bm{p}, \bm{n})}{N}} \frac{Q^{-1}(\epsilon)}{\ln 2}, \label{eq:uplink_rate}
\end{split}
\end{equation}
where $B$ represents the system bandwidth, $\epsilon$ characterizes the likelihood of decoding errors, $\gamma_{m\tilde{\Phi}_L}(p,n)$ stands for the Signal-to-Noise Ratio (SNR) observed by the $m$-th UE, $Q^{-1}(.)$ is the reverse function of $Q(x) = \frac{1}{\sqrt{2\pi}} \int_x^{\infty} e^{-t^2/2} \,dt$, and $V_{m\tilde{\Phi}_L}$ is the channel dispersion given as $V_{m\tilde{\Phi}_L}(p,n) = 1 - \left[1 + \gamma_{m\tilde{\Phi}_L}(\bm{p}, \bm{n})\right]^{-2}$.
This equation computes the uplink rate for the chosen destination, accounting for channel characteristics, bandwidth allocation, transmit power, and more.

Subsequently, the uplink wireless transmission latency from $m$-th UE to the $\tilde{\Phi}_L$-th C-MEC resource can be expressed as:
\begin{equation}
    T_{m\tilde{\Phi}_L}^{\text{CO}}(\bm{p}, \bm{n},\tilde{\Phi}_L) = \max_{\forall\Phi_L} \left\{\frac{\Phi_L I_m}{\aleph_{m\Phi_L}(\bm{p}, \bm{n})}\right\}. \label{eq:uplink_latency}
\end{equation}

\subsection{Computation Model}
In the computation model, each UE generates granular computation task $J_m$ in which a portion can be executed by the CNs and another portion at the ES. The model is defined as follows:

\subsubsection{COIN Node Processing}
For the COIN node, the task $J_m$ portion $\lambda_{mk}$ is executed by the CNs with the estimated processing rate $f_m^{cn}$. Consequently, the estimated CN execution latency is given as:
\begin{equation}
    \tilde{T}_{mk}^{cn}(\lambda_{mk},f_m^{cn}) = \max_{\forall k \in K} \left\{\frac{\lambda_{mk} C_m}{f_m^{cn}}\right\}. \label{eq:coin_processing_time}
\end{equation}

Assuming we can pre-determine the discrepancy between the actual $k$-th CN and its DT, we can estimate the gap in computing latency between real-world performance and DT predictions as follows:
\begin{equation}
    \Delta T_{mk}^{cn}(\lambda_{mk},f_m^{cn}) = \frac{\lambda_{mk} C_m \tilde{f}_{mk}^{cn}}{f_m^{cn} (f_m^{cn} - \tilde{f}_{mk}^{cn})}. \label{eq:delta_latency}
\end{equation}

Thus, the actual CNs processing time is $ T_{mk}^{cn} = \Delta T_{mk}^{cn} + \tilde{T}_{mk}^{cn}$. The total latency, including transmission and computing latency is given as 
\begin{equation}
T_m^{cnT} = T_{mk}^{cn} + T_{m\Phi_L}^{CO}
\end{equation}

\subsubsection{MEC Processing}
The task $J_m$ portion $\aleph_m$ executed by the ES with the estimated processing rate $f_m^{em}$ incurs the following latency:
\begin{equation}
\tilde{T}_m^{em}(\aleph_m,f_m^{em}) = \frac{\aleph_m C_m}{f_m^{em}}. \label{eq:mec_processing_time}
\end{equation}

The latency gap $\Delta T_m^{em}$ between the real latency and the DT is estimated as
\begin{equation}
\Delta T_m^{em}(\mathcal{N}_m,f_m^{em}) = \frac{\mathcal{N}_m C_m \tilde{f}_m^{em}}{f_m^{em}(\tilde{f}_m^{em}-f_m^{em})} \quad
\end{equation}

Consequently, the actual latency for task execution at $
T_m^{em} = \Delta T_m^{em} + \tilde{T}_m^{em} \quad $. The total delay at MEC is thus;
\begin{equation}
T_m^{emT} = T_{m\Phi_L}^{CO} + T_m^{em}
\end{equation}

\subsubsection{ Latency model}
The total end-to-end (e2e) DT latency within the system includes the UEs processing latency, task offloading transmission latency, and the ES processing latency. Thus, the e2e DT latency is expressed as
\( T_m^{e2e} = T_m^{kcn} + T_{m\Phi_L}^{CO} + T_m^{em} = \max_{\forall k\in K}\left\{\frac{\mathcal{L}_m C_m}{f_m^{kcn}-\tilde{f}_m^{kcn}}\right\} + \max_{\forall\Phi_L}\left\{\frac{\Phi_L I_m}{\aleph_{m\Phi_L}(p,n)}\right\} + \frac{\mathcal{N}_m C_m}{f_m^{em}-\tilde{f}_m^{em}} \).
\vspace{-5.7mm}
\subsection{Problem formulation}
\vspace{-4.805mm}
Let $S_m=\{s_{m0},s_{m1},s_{m2},\ldots,s_{mK}\mid s_{mj} \in \{0,1\}\}$ denote the offloading strategies for UE $m$. The offloading strategy profile of all UEs is denoted as $s=\{s_m\mid s_m \in S_m, m \in \mathcal{M}\}$, where $s_m=s_{mj}=1$ suggests that UE $m$ accomplishes its task via decision $j$, otherwise $s_m=s_{m0}=0$. $s_{m0}$ indicates the decision variable for task execution at the ES while $s_{mk}$ are executed at the CN $k$ node.

From the UEs perspective, we define the UE $m$ utility as the difference between the reduced latency due to offloading and the computational cost as follows \cite{pham2022partial}:
\begin{equation}
U_m = \sum_{j\in \mathcal{K}\cup\text{\{0\}}} s_{mj}[g_t(T_m^{em}-T_m^{e2e})-p_j\Phi_j C_m] \quad
\end{equation}

where $g_t$ is the unit gain latency reduction, and $p_j$ is proportional to computing capacity, indicating offloading cost per workload at node $j$.

Our primary objective, denoted by $\mathcal{P}$, is to maximize the system utility by minimizing the overall system latency, considering the optimal offloading ratio and resource allocation. This is formalized as follows:

\begin{align}
    \mathcal{P}: & \max_{(s,\Phi,\beta)} \sum_{m\in \mathcal{M}} U_m & \label{eq:objective} \\
    \text{s.t.} & & \nonumber \\
    & \sum_{j\in K\cup\{0\}} s_{mj} \leq 1, \quad \forall m \in \mathcal{M} \tag{10a} \label{eq:constraint_10a} \\
    & \sum_{m\in M} s_{mj} \leq 1, \quad \forall j \in K\cup\{0\} \tag{10b} \label{eq:constraint_10b} \\
    & s_{mj} T_m^{e2e} \leq T_{\text{max}}, \quad \forall m \in \mathcal{M}, j \in K\cup\{0\} \tag{10c} \label{eq:constraint_10c} \\
    & \sum_{m\in M} \beta_m \leq 1 \tag{10d} \label{eq:constraint_10d} \\
    &  s_{mj} \in \{0,1\}, 0 \leq \Phi, \beta \leq 1, \quad \forall m\in \mathcal{M}, & \nonumber\\
    & j\in K\cup\{0\} \tag{10e} \label{eq:constraint_10e}
\end{align}

Constraint (10a) suggests that each task is partially offloaded to at most one computing node. (10b) represents the subsystem to COIN node association constraints. (10c) enforces the latency requirement. (10d) guarantees that allocated computing resources are within the limit of the CN capacity. (10e) denotes the constraints of optimizing variables.

\section{Proposed solution}
The objective function exhibits non-convex characteristics due to partial offloading decision variables and non-linear relationships. It is intractable to solve directly since it involves PCO in C-MEC cyber twin across different time slots and lacks UE request transition probabilities. To address this complexity, we decompose the DT problem into two subproblems: partial offloading decision problem and offloading ratio and resource allocation problem.

\subsection{Multi-user Computation Offloading Game}
The multi-user computation offloading game can be defined as $G=\{M, (S_m)_{m\in \mathcal{M}}, (U_m)_{m\in \mathcal{M}}$, where $S_m$ is the set of offloading strategies for UE $m$, and $U_m(s_m,\bm{s}_{(-m)})$ is the utility function taking into account the set of offloading strategies. Here, $s_{-m}=(s_1,\ldots,s_{(m-1)},s_{(m+1)},\ldots,s_M)$ represents the offloading strategies of all UEs except the $m$th. Each UE selects the most advantageous strategy that enhances its individual utility. The game is considered to achieve a state of Nash Equilibrium (NE) when no UE can further improve its utility by altering its offloading choice. \\
\textit{Definition 1:} A strategy $s^*=(s_1^*,s_2^*,\ldots,s_M^*)$ is the NE of the game $G$ if it adheres to

\begin{equation}
\begin{gathered}
U_m(s_m^*,\bm{s}_{(-m)}^*) \geq U_m(s_m,\bm{s}_{(-m)}^*), \\
\forall m\in M, \forall s_m\in S_m.
\end{gathered}
\end{equation}

Based on \cite{pham2022partial}, the game $G$ is an exact potential game (EPG) by formulating the potential function as follows:

\begin{equation}
\begin{aligned}
\phi(s) = &s_{m0} \sum_{m\in M}\left[ R_{m0} + (1-s_{m0})\left(\sum_{j\in K}s_{mj}R_{mj} \right. \right.\\
&\left. \left. + \sum_{m'\neq m} R_{m'0}\right)\right] \quad 
\end{aligned}
\end{equation}
where $R_{mj} = g_t(T_m^{em} - T_m^{kcn}) - p_j \Phi_j C_m$.
For ease of proof, the expression $\phi(s_m,s_{(-m)})$ is given as:
\begin{equation}
\begin{aligned}
\phi(s_m,s_{(-m)}) = \, 
s_{m0} \sum_{m\in M}\bigg[ R_{m0} + (1-s_{m0}) \\ 
\bigg(\sum_{j\in K}s_{mj}R_{mj}
 + \sum_{m'\neq m} R_{m'0}\bigg)\bigg]
\end{aligned}
\end{equation}

\textit{Remark 1:} The game $G$ with the potential function $\phi(s)$ is an EPG and capable of reaching an NE in a finite number of iterations.

\subsection{DDQN for optimal offloading ratio and resource allocation}
For maximizing the utility, the joint optimization of offloading ratio and resource allocation (ORRA) problem can be reformulated as follows:

\begin{align}
    & \mathcal{P}_1: & & \min_{\Phi,\beta} \sum_{m\in M} p_j \Phi_j C_m - (T_m^{em} - T_m^{e2e}) \\
    & \text{s.t.} & & \nonumber \\
    & & & T_m^{e2e} \leq T_{\text{max}}, \tag{14a} \label{eq:constraint_24a} \\
    & & & \sum_{m\in M} \beta_m \leq 1, \tag{14b} \label{eq:constraint_24b} \\
    & & & 0 \leq \Phi, \beta \leq 1, \quad \forall m\in M. \tag{14c} \label{eq:constraint_24c}
\end{align}

For any time slot $(t+1)$ given the user offloading request $\mu_{(t+1)}$, the optimal offloading ratio $\Phi_{(t+1)}$ and resource allocation $\beta_{(t+1)}$ can be solved. However, the $\mu_{(t+1)}$ is unknown due to unknown user request transition probabilities. The DDQN is employed to capture users' request model and predict the optimal task offloading ratio and corresponding resource allocation of time slot $(t+1)$ based on the system state at slot $t$.

We formulate the $\mathcal{P}_1$ as a Markov Decision Process (MDP) and elaborate on the state, action, and reward as follows:\\
\textbf{State:} The user request state at time slot $t$ is denoted as $S_t = \mu_t \in (F+1)^M$, where $F$ is the number of tasks.

\textbf{Action:} The action at time slot $t$ is the offloading ratio and resource allocation $A_t = \Phi_{t+1}, \beta_{t+1} \in [0,1]^M$.

\textbf{Reward:} The reward at time $t$ is defined as the utility savings in time $(t+1)$, denoted as $R_{t+1}$. This saving is calculated as the difference between the utility derived from the optimal partial offloading ratio and resource allocation, and that from full offloading ratio and resource allocation at the same time.

\subsection{Game-Theoretic Offloading Framework (GTOF)}

The  Game-Theoretic Offloading Framework (GTOF) (Algorithm 1) solve the partial computation offloading decision problem ($\mathcal{P}$) using the future optimal ORRA Problem $(\mathcal{P}_1)$ for efficient computation offloading. The Base Station (BS) acts as the central hub in its operation, assimilating real-time data like connection statuses and UE strategies. Initially, Service Modules (SMs) lean towards MEC offloading. However, as iterations progress, each UE refines its offloading strategy based on feedback from the BS. This iterative exchange continues until the UEs seek no further updates, indicating a Nash Equilibrium. The computational complexity of GTOF is represented as $O(C1 \times N )$, where $C1$ is the iteration count for the DDQN.

\begin{algorithm}
    \caption{Multiuser PCO}
    \label{alg:multiuser_pco}
    \begin{algorithmic}[1]
        \State \textbf{Initialization:} Each UE $k \in K$ initializes its PCO decision towards offloading to the MEC server.
        \For{decision slot $t$:}
            \For{each UE $k \in K$:}
                \State Obtain real-time system environment from the Base Station (BS).
                \If{there exist available Digital Twins}
                    \State Determine $U_{ij}$ according to equations pertaining to the OORRA Problem ($\mathcal{P}_1$).
                \EndIf
                \State Obtain the best optimal strategy such that $U_m(s_m',s_{-m})=\arg\min_{S_{(k,t)}} U_{mj}$
                \If{$U_m(s_m',s_{-m}) > U_m(s_m,s_{-m})$}
                    \State The updated PCO strategy, $S_{(k,t)} = S_{(k,t)}^*$, is sent to the BS and stored into $M(t)$.
                \Else
                    \State Retain the previous strategy $s_m$.
                \EndIf
            \EndFor
            \If{$M(t) \neq \emptyset$}
                \State Each UE in $M(t)$ vies for the next update opportunity.
                \If{UE $i$ wins}
                    \State Broadcast the update to all UE: $s_m(t) = s_m'$.
                \Else
                    \State Retain the previous strategy $s_m$.
                \EndIf
            \EndIf
        \EndFor
        \Repeat
            \Until an END message is received
        \State \textbf{Return:} The optimal offloading strategies $s^*$.
    \end{algorithmic}
\end{algorithm}

\section{Numerical Results}

This section presents the numerical results and analysis of our simulation to evaluate the performance of our proposal. We considered C-MEC networks where UEs are randomly distributed in a $200 \, \text{m} \times 200 \, \text{m}$ area with $[4,12]$ UEs, $[1,10]$ COIN nodes, and an ES server. The large-scale fading from the $m$-th User Equipment (UE) to the $k$-th Access Point (AP) is modeled as $g_m = 10^{\left(\frac{PL(d_{mk})}{10}\right)}$ with path loss $PL(d_{mk}) = -35.3 - 37.6 \log_{10}(d_{mk}) = -35.3 - 37.6$ \cite{van2022urllc}; Noise spectral density is set to $-174 \, \text{dBm/Hz}$ \cite{nasir2020resource}, bandwidth to $10 \, \text{MB}$ and URLLC decoding error probability is $\epsilon = 10^{-9}$. See Table I for additional parameters.

\begin{table}[ht]
    \centering
    \caption{Simulation Parameters}
    \begin{tabular}{ll}
        \hline
        \textbf{Parameters} & \textbf{Value} \\
        \hline
        $I_m$  & [1, 10] MB\\
         $C_m$  , and  $T_m^{max}$ & [0.001, 0.1] GHz; 15 ms\\
        MEC computing capacity & 30 GHz \cite{pham2022partial} \\
        COIN node capacity & [1, 10] GHz \cite{lia2022in} \\
        Transmission power of UEs &   \\
        Unit gain of latency reduction & 2.5\\
        Offloading cost per workload & $0.1  \times 10\text{GHz}$\\
        Experience Memory & $10000$\\
        Discount factor & 0.9\\
        \hline
    \end{tabular}
\end{table}

To verify the effectiveness of the proposed method, we evaluate our approach against the following baselines:

\begin{itemize}
  \item \textit{Our Scheme (DDQN-EPG):} Employing DDQN to predict future optimal ORRA in a game theoretic framework based on EPG to maximize user utility in a C-MEC network.
  
  \item \textit{EPG with Random ORRA (EPG-Rand):} This strategy is based on randomly predicted future offloading ratio and resource allocation. This baseline gives insight into the overall future system performance when DDQN is not applied.
  
  \item \textit{MEC:} This is the Conventional MEC network with no COIN capabilities enabled, in which UEs can perform the task locally or offload it to the MEC. This baseline allows for a direct comparison between the proposed COIN approach and the standard MEC baseline, highlighting the performance improvement
\end{itemize}

\begin{figure}[ht]
  \centering
  \begin{subfigure}{0.48\columnwidth}
    \includegraphics[width=\linewidth]{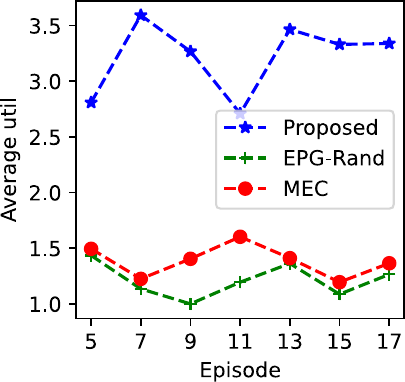}
    \caption{}
    \label{fig:subfig1a}
  \end{subfigure}
  \hfill
  \begin{subfigure}{0.48\columnwidth}
    \includegraphics[width=\linewidth]{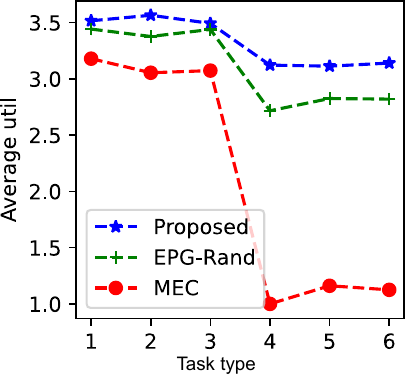} 
    \caption{}
    \label{fig:subfig1b}
  \end{subfigure}
  \caption{Performance evaluation based on Experimental Parameters on the System Model (a) Episode (b) Task input type}
  \label{Performance evaluation based on Experimental Parameters on the System Model (a) training episode (b) Task input type}
\end{figure}

\begin{figure}[ht]
  \centering
  \begin{subfigure}{0.48\columnwidth}
    \includegraphics[width=\linewidth]{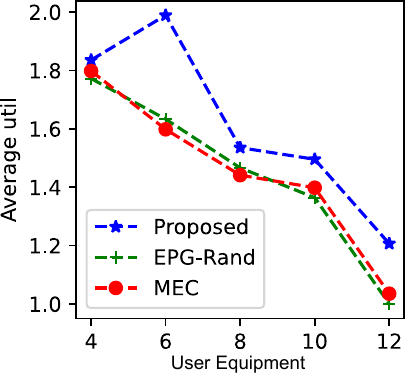}
    \caption{}
    \label{fig:subfig2a}
  \end{subfigure}
  \hfill
  \begin{subfigure}{0.48\columnwidth}
    \includegraphics[width=\linewidth]{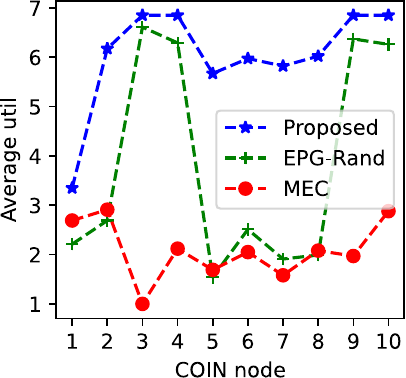} 
    \caption{}
    \label{fig:subfig2b}
  \end{subfigure}
  \caption{Performance evaluation based on different numbers of users/COIN node (a) UE number (b) COIN nodes}
  \label{Performance evaluation based different numbers of users/COIN node (a) UEs number (b) COIN nodes}
\end{figure}
\vspace{0.2em}
In order to ensure a fair performance comparison, we conducted a comprehensive analysis of various aspects. Our evaluation involved comparing the average system utility across training episodes against benchmark scenarios. Except for episode 5, our model consistently achieved the highest average system utility as shown in Fig.~\ref{fig:subfig1a}. In a few episodes of 21, our scheme attains 20\% utility over the baseline, demonstrating effective offloading ratio and resource management via the DDQN.

Furthermore, the proposed system model's effectiveness is evaluated by investigating the influence of computing task types: data-intensive and compute-intensive types. For data-intensive tasks (Tasks 1 to 3), the input size ($I_m$) and required CPU cycles ($C_m$) of the tasks are uniformly and randomly generated from the ranges {[10-20] MB, [0.1-0.5] GB}, respectively. In the compute-intensive task type, $I_m$ and $C_m$ are uniformly and randomly generated from the ranges {[1-5] MB, [1-2] GB}, respectively. Considering the average system utility, our model consistently outperformed others, with an increase of 43.0\% to 87.9\% for data-intensive tasks (1–3) and 36.2\% to 87.7\% for compute-intensive tasks (4–6) compared to the second-best MEC model, as shown in Fig.~\ref{fig:subfig1b}.

Next, we evaluate the performance of the proposed model by investigating the impact of varying UEs and the COIN node numbers. For various UE numbers, our proposed method consistently demonstrates superior utility-effectiveness in Fig.~\ref{fig:subfig2a}, showcasing a remarkable 47\% increment in utility compared to the baselines. Notably, the increase in UE beyond 6 resulted in an overall reduction in the average system utility. Examining the COIN node number in Fig.~\ref{fig:subfig2b}, our proposed model excels with a significant 64\% improvement over the baselines between 5 to 8 COIN nodes. Although the EPG-Rand significantly improves beyond 8 COIN nodes, our model maintains improved performance, underscoring our approach's ability to enhance the OPG algorithm, making it more efficient in increasing COIN-enabled nodes.

\section{Conclusion}
This paper explores a digital twin (DT) scheme for collaborative task offloading in a multiuser scenario involving user equipment (UE) and COIN-aided nodes or Mobile-Access Edge Computing (MEC). The approach formulates the problem as a unified decision-making process for task offloading, offloading ratio and resource allocation. Leveraging game theory, a low-complexity distributed offloading scheme is devised, treating the task decision problem as an exact potential game. The inclusion of Double Deep Q-Network (DDQN) allows proactive prediction of optimal offloading ratios and resource allocations. The DT-based emulation of C-MEC provides insights into processing rates, enabling efficient computing resource allocation and reduced processing latency. Subsequent research can explore the scheme's performance under varied input parameter sizes, probabilistic user requests and its impact on energy consumption.

\section*{Acknowledgment}
This work was partly supported by the Innovative Human Resource Development for Local Intellectualization program through the Institute of Information \& Communications Technology Planning \& Evaluation(IITP) grant funded by the Korean government(MSIT) (IITP-2024-RS-2022-00156287, 50); and in part by Institute of Information \& communications Technology Planning \& Evaluation (IITP) grant funded by the Korea government(MSIT) (No.2021-0-02068, Artificial Intelligence Innovation Hub, 50)

\bibliographystyle{elsarticle-num}
\bibliography{bibliography}

\end{document}